\documentclass[letterpaper]{article} 
\usepackage{aaai24}  
\usepackage{times}  
\usepackage{helvet}  
\usepackage{courier}  
\usepackage[hyphens]{url}  
\usepackage{graphicx} 
\usepackage{natbib}  
\usepackage{caption} 
\usepackage{algorithm}
\usepackage{algorithmicx}
\usepackage{algpseudocode}

\usepackage{diagbox}
\usepackage{booktabs}       
\usepackage{amsfonts}       
\usepackage{nicefrac}       
\usepackage{microtype}      
\usepackage{xcolor}         
\usepackage{subfigure}
\usepackage{bbm}
\usepackage{amssymb}

\usepackage{graphicx}
\usepackage{amsmath}
\usepackage[utf8]{inputenc} 
\usepackage{url}            
\usepackage{array}
\usepackage{bm}
\usepackage{dsfont}
\usepackage{amsthm}

\newtheorem{mydef}{Definition}

\usepackage{newfloat}
\usepackage{listings}
\DeclareCaptionStyle{ruled}{labelfont=normalfont,labelsep=colon,strut=off} 
\floatstyle{ruled}
\newfloat{listing}{tb}{lst}{}
\floatname{listing}{Listing}
\title{TAPE: Leveraging Agent Topology for Cooperative Multi-Agent Policy Gradient}
\author{
    Xingzhou Lou\textsuperscript{\rm 1,2}\thanks{Work done while visiting King's College London}, Junge Zhang\textsuperscript{\rm 1,2}\thanks{Correspondence}, Timothy J. Norman\textsuperscript{\rm 3}, Kaiqi Huang\textsuperscript{\rm 1,2}, Yali Du\textsuperscript{\rm 4}
}
\affiliations{
    \textsuperscript{\rm 1}School of Artificial Intelligence, University of Chinese Academy of Sciences\\
    \textsuperscript{\rm 2}Institute of Automation, Chinese Academy of Sciences\\
    \textsuperscript{\rm 3}University of Southampton\\
    \textsuperscript{\rm 4}King's College London\\
    louxingzhou2020@ia.ac.cn, jgzhang@nlpr.ia.ac.cn, t.j.norman@soton.ac.uk\\
    kqhuang@nlpr.ia.ac.cn, yali.du@kcl.ac.uk
}

\usepackage{bibentry}

\begin{document}

\maketitle

\begin{abstract}
Multi-Agent Policy Gradient (MAPG) has made significant progress in recent years. However, centralized critics in state-of-the-art MAPG methods still face the centralized-decentralized mismatch (CDM) issue, which means sub-optimal actions by some agents will affect other agent's policy learning. While using individual critics for policy updates can avoid this issue, they severely limit cooperation among agents. To address this issue, we propose an agent topology framework, which decides whether other agents should be considered in policy gradient and achieves compromise between facilitating cooperation and alleviating the CDM issue. The agent topology allows agents to use coalition utility as learning objective instead of global utility by centralized critics or local utility by individual critics. To constitute the agent topology, various models are studied. We propose Topology-based multi-Agent Policy gradiEnt (TAPE) for both stochastic and deterministic MAPG methods. We prove the policy improvement theorem for stochastic TAPE and give a theoretical explanation for the improved cooperation among agents. Experiment results on several benchmarks show the agent topology is able to facilitate agent cooperation and alleviate CDM issue respectively to improve performance of TAPE. Finally, multiple ablation studies and a heuristic graph search algorithm are devised to show the efficacy of the agent topology.
\end{abstract}

\section{Introduction}
Recent years has witnessed dramatic progress of reinforcement learning (RL) and multi-agent reinforcement learning (MARL) in real life applications, such as unmanned vehicles \cite{liu2022reinforcement}, traffic signal control \cite{noaeen2022reinforcement} and on-demand delivery \cite{wang2023cross}. Taking advantage of the centralized training decentralized execution (CTDE) \cite{oliehoek2008optimal,kraemer2016multi} paradigm, current cooperative MARL methods \cite{du2023review,wang2020qplex,wang2020rode,peng2021facmac,zhang2021fop,zhou2022pac} adopt value function factorization or a centralized critic to provide centralized learning signals to promote cooperation and achieve implicit or explicit credit assignment. 
Multi-agent policy gradient (MAPG) \cite{lowe2017multi,foerster2018counterfactual,zhou2020learning,zhang2021fop,zhou2022pac,du2019liir} applies RL policy gradient techniques \cite{sutton2018reinforcement,silver2014deterministic,lillicrap2015continuous} to the multi-agent context. In CTDE, MAPG methods adopt centralized critics or value-mixing networks \cite{rashid2020monotonic,rashid2020weighted,wang2020qplex} for individual critics so that agents can directly update their policies to maximize the global $Q$ value $Q^{\bm{\pi}}_{tot}$ in their policy gradient. As a result, agents cooperate more effectively and obtain better expected team rewards.

The centralized critic approach has an inherent problem known as centralized-decentralized mismatch (CDM) \cite{wang2020off,chen2022learning}. The CDM issue refers to sub-optimal, or explorative actions of some agents negatively affecting policy learning of others, causing catastrophic miscoordination. The CDM issue arises because sub-optimal or explorative actions may lead to a small or negative centralized global $Q$ value $Q^{\bm{\pi}}_{tot}$, even if other agents take good or optimal actions. In turn, the small $Q^{\bm{\pi}}_{tot}$ will make the other agents mistake their good actions as bad ones and interrupt their policy learning. The Decomposed Off-Policy (DOP) approach \cite{wang2020off} deals with sub-optimal actions of other agents by linearly decomposed individual critics, which ignore the other agents' actions in the policy gradient. But the use of individual critics severely limits agent cooperation.

We give an example to illustrate the issue of learning with centralised critics and individual critics respectively. Consider an one-step matrix game with two agents $A$, $B$ where each agent has two actions $a_0,a_1$. Reward $R(a_0,a_0)=2,R(a_0,a_1)=-4,R(a_1,a_0)=-1$ and $R(a_1,a_1)=0$.
Assume agent $A$ has a near-optimal policy with probability $\epsilon$ choosing non-optimal action $a_1$ and is using the COMA centralized critic \cite{foerster2018counterfactual} for policy learning. If agent $A$ takes optimal action $a_0$ and $B$ takes the non-optimal action $a_1$, agent $A$'s counterfactual advantage $Adv_A(a_0,a_1)=Q_{tot}^{\bm{\pi}}(a_0,a_1)-\left[\left(1-\epsilon\right)Q_{tot}^{\bm{\pi}}(a_0,a_1)+\epsilon Q_{tot}^{\bm{\pi}}(a_1,a_1)\right]=-4\epsilon<0$, which means agent $A$ will mistakenly think $a_0$ as a bad action. Consequently, the sub-optimal action of agent $B$ causes agent $A$ to decrease the probability of taking optimal action $a_0$ and deviate from the optimal policy. Similar problems will occur with other centralized critics. If we employ individual critics, however, cooperation will be limited. Assume both agents' policies are initialized as random policies and learning with individual critics. For agent $A$, $Q_A(a_0)=\mathbb{E}_{a_B\sim\pi_B}[Q^{\bm{\pi}}_{tot}(a_0,a_B)]=0.5\times2-0.5\times4=-1$. Similarly, we can get $Q_A(a_1)=-0.5,\ Q_B(a_0)=0.5,\ Q_B(a_1)=-2$. The post-update joint-policy will be $(a_1,a_0)$ and receive reward $-1$, which is clearly sub-optimal.

In this paper, we aims to alleviate the CDM issue without hindering agent's cooperation capacity by proposing an agent topology framework to describe the relationships between agents' policy updates. Under the agent topology framework, agents connected in the topology consider and maximize each other's utilities. Thus, the shared objective makes each individual agent forms a coalition with its connected neighbors. Agents only consider the utilities of agents in the same coalition, facilitating in-coalition cooperation and avoiding influence of out-of-coalition agents. Based on the agent topology, we propose \textbf{T}opology-based multi-\textbf{A}gent \textbf{P}olicy gradi\textbf{E}nt (TAPE) for both stochastic and deterministic MAPG, where the agent topology can alleviate the bad influence of other agents' sub-optimality without hindering cooperation among agents. Theoretically, we prove the policy improvement theorem for stochastic TAPE and give a theoretical explanation for improved cooperation by exploiting agent topology from the perspective of parameter-space exploration.

Empirically, we use three prevalent random graph models \cite{erdHos1960evolution,watts1998collective,albert2002statistical} to constitute the agent topology. Results show that the Erdős–Rényi (ER) model \cite{erdHos1960evolution} is able to generate the most diverse topologies. With diverse coalitions, agents are able to explore different cooperation patterns and achieve strong cooperation performance. Evaluation results on a matrix game, Level-based foraging \cite{papoudakis2021benchmarking} and SMAC \cite{samvelyan19smac} show that TAPE outperforms all baselines and the agent topology is able to improve base methods' performance by both facilitating cooperation among agents and alleviating the CDM issue. Moreover, to show the efficacy of the agent topology, we conduct multiple studies and devise a heuristic graph search algorithm.

Contributions of this paper are three-fold: Firstly, We propose an agent topology framework and Topology-based multi-Agent Policy gradiEnt (TAPE) to achieve compromise between facilitating cooperation and alleviating CDM issue; Secondly, we theoretically establish policy improvement theorem for stochastic TAPE and elaborate the cause for improved cooperation by agent topology; Finally, empirical results demonstrate that the agent topology is able to alleviate the CDM issue without hindering cooperation among agents, resulting in strong performance of TAPE.

\section{Preliminaries}\label{bg}
The cooperative multi-agent task in this paper is modelled as \textbf{Decentralized Partially Observable Markov Decision Process} (Dec-POMDP) \cite{oliehoek2016concise}. A Dec-POMDP is a tuple $G=\left\langle I, S,\mathcal{A},P,r,\mathcal{O},O,n,\gamma\right\rangle$, where $I=\{1,..,n\}$ is a finite set of $n$ agents, $S$ is the state space, $\mathcal{A}$ is the agent action space and $\gamma$ is a discount factor. At each timestep, every agent $i \in I$ picks an action $a_i \in \mathcal{A}$ to form the joint-action $\mathbf{a}\in\mathbf{A}= \mathcal{A}^n$ to interact with the environment. Then a state transition will occur according to a state transition function $P(s'|s,\mathbf{a}):S\times\mathbf{A}\times S\rightarrow [0,1]$. All agents will receive a shared reward by the reward function $r(s,\mathbf{a}):S\times \mathbf{A}\rightarrow \mathbb{R}$. During execution, every agent draws a local observation $o\in \mathcal{O}$ by an observation function $O(s,a):S\times A\rightarrow \mathcal{O}$. Every agent stores an observation-action history $\tau^a\in T=(\mathcal{O}\times \mathcal{A})$, based on which agent $i$ derives a policy $\pi_i(a_i|\tau_i)$. The joint policy $\bm{\pi}=\{\pi_1,..,\pi_n\}$ consists of policies of all agents. The global $Q$ value function $Q^{\boldsymbol{\pi}}_{tot}(s,\mathbf{a})=\mathbb{E}_{\bm{\pi}}[\sum_{i=0}\gamma^ir_{t+i}|s_t=s,\mathbf{a}_t=\mathbf{a}]$ is the expectation of discounted future reward summed over the joint-policy $\boldsymbol{\pi}$.

The policy gradient in stochastic MAPG method DOP is: $g=\mathbb{E}_{\bm{\pi}}\left[\sum_ik_i(s)\nabla_{\theta_i}\log\pi_i(a_i|\tau_i)Q_i^{\phi_i}(s,a_i)\right]$, where $k_i\geq0$ is the positive coefficient provided by the mixing network, and the policy gradient in deterministic MAPG methods is $g=\mathbb{E}_{\mathcal{D}}\left[\sum_i \nabla_{\theta_i}\pi_i(\tau_i)\nabla_{a_i}Q^{\bm{\pi}}_{tot}(s,\bm{a})|_{a_i=\pi_i(\tau_i)}\right]$, where $Q^{\bm{\pi}}_{tot}$ is the centralized critic and $\pi_i$ is the policy of agent $i$ parameterized $\theta_i$.

\section{Related Work}\label{related_work}

\textbf{Multi-Agent Policy Gradient} The policy gradient in \textbf{stochastic MAPG} methods has the form $\mathbb{E}_{\bm{\pi}}\left[\sum_i\nabla_{\theta_i}\log\pi_i(a_i|\tau_i)G_i\right]$ \cite{foerster2018counterfactual,wang2020off,lou2023leveraging,chen2022learning}, where objective $G_i$ varies across different methods, such as counterfactual advantage \cite{foerster2018counterfactual} and polarized joint-action value \cite{chen2022learning}. The objective in DOP is individual aristocratic utility \cite{wolpert2001optimal}, which ignores other agents' utilities to avoid the CDM issue, but the cooperation is also limited by this objective. It is worth noting that polarized joint-action value \cite{chen2022learning} also aims to address the CDM issue, but it only applies to stochastic MAPG methods, and the polarized global $Q$ value can be very unstable. \textbf{Deterministic MAPG} methods use gradient ascent to directly maximize the centralized global $Q$ value $Q^{\bm{\pi}}_{tot}$. Lowe \textit{et al.} \cite{lowe2017multi} model the global $Q$ value with a centralized critic. Current deterministic MAPG methods \cite{zhang2021fop,peng2021facmac,zhou2022pac} adopt value factorization to mix individual $Q$ values to get $Q^{\bm{\pi}}_{tot}$. As the global $Q$ value is determined by the centralized critic for all agents, sub-optimal actions of one agent will easily influence all others.

\noindent\textbf{Topology in Reinforcement Learning} Adjodah \textit{et al.} \cite{adjodah2019leveraging} discuss the communication topology issue in parallel-running RL algorithms such as A3C \cite{mnih2016asynchronous}. Results show that the centralized learner implicitly yields a fully-connected communication topology among parallel workers, which will harm their performance. In MARL with decentralized training, communication topology is adopted to enable inter-agent communication among networked agents \cite{zhang2018fully,wang2019learning,konan2022iterated,du2021flowcomm}. The communication topology allows agent to share local information with each other during both training and execution and even achieve local consensus, which further leads to better cooperation performance. In MARL with centralized training, deep coordination graph (DCG) \cite{bohmer2020deep} factorizes the joint value function according to a coordination graph to achieve a trade-off between representational capacity and generalization. Deep implicit coordination graph \cite{li2020deep} allows to infer the coordination graph dynamically by agent interactions instead of domain expertise in DCG. Ruan \textit{et al.} \cite{ruan2022gcs} learn an action coordination
graph to represents agents' decision dependency, which further coordinates the dependent behaviors among agents.

\section{Topology-based Multi-Agent Policy Gradient}
In this section, we propose \textbf{T}opology-based multi-\textbf{A}gent \textbf{P}olicy gradi\textbf{E}nt (TAPE), which exploits the agent topology for both stochastic and deterministic MAPG. This use of the agent topology provides a compromise between facilitating cooperation and alleviating CDM. The primary purpose of the agent topology is to indicate relationships between agents' policy updates, so we focus on policy gradients of TAPE here and cover the remainder in supplementary material. First, we will define the agent topology.

The agent topology describes how agents should consider others' utility during policy updates. Each agent is a vertex $v\in \mathcal{V}$ and $\mathcal{E}$ is the set of edges. For a given topology, $(\mathcal{V},\mathcal{E})$, if  $e_{ij}\in \mathcal{E}$, the source agent $i$ should consider the utility of the destination agent $j$ in its policy gradient. The only constraint we place on a topology is that $\forall\ i,\ e_{ii}\in\mathcal{E}$, because agents should at least consider their own utility in the policy gradient. The topology captures the relationships between agents' policy updates, not their communication network at test time \cite{foerster2016learning,das2019tarmac,wang2019learning,ding2020learning}. Connected agents consider and maximize each other's utilities together. Thus, the shared objective makes each individual agent form a coalition with the connected neighbors. We use the adjacency matrix $E$ to refer the agent topology in what follows.

In our agent topology framework, DOP \cite{wang2020off} (policy gradient given in section \ref{bg}) and other independent learning algorithms' has an edgeless agent topology. The adjacency matrix is the identity matrix and no edge exists in the topology. With no coalition, DOP agent will only maximize its own individual utility $Q_i$, and hence is poor at cooperation. Although DOP adopts a mixing network for the individual utilities to enhance cooperation, an agent's ability to cooperate is still limited, which we will empirically show in the matrix game experiments. Methods with centralized critic such as COMA \cite{foerster2018counterfactual}, FACMAC \cite{peng2021facmac} and PAC \cite{zhou2022pac} yields the fully-connected agent topology. In these methods, there is only one coalition with all of the agents in it (all edges exist in the topology), and all agents update their policies based on the centralized critic. Consequently, they suffer from the CDM issue severely, because the influence of an agent's sub-optimal behavior will spread to the entire multi-agent system.

\subsection{Stochastic TAPE}
Instead of global centralized critic \cite{foerster2018counterfactual}, we use the agent topology to aggregate individual utilities and critics to facilitate cooperation among agents for stochastic MAPG \cite{wang2020off}. To this end, a new learning objective \emph{Coalition Utility} for the policy gradient is defined as below.

\begin{mydef}[Coalition Utility]  
Coalition Utility $\bm{U}_i$ for agent $i$ is the summation of individual utility $U_j$ of connected agent $j$ in agent topology $E$, i.e. $\bm{U}_i=\sum_{j=1}^nE_{ij}U_j$,
where $U_j(s,a_j)=Q^\phi_{tot}(s,\bm{a})-\sum_{a_j'}\pi_j({a_j'}|\tau_j)Q^\phi_{tot}(s,({a_j'},\bm{a}_{-j}))$.
   
\end{mydef}

$U_j$ is the aristocrat utility from \cite{wang2020off,wolpert2001optimal}. 
$E_{ij}=1$ only if agent $j$ is connected to agent $i$ in $E$ and $Q^\phi_{tot}$ is the global $Q$ value function. Coalition utility only depends on in-coalition agents because if agent $j$ is not in agent $i$'s coalition, $E_{ij}=0$. With the coalition utility, we propose \textbf{stochastic TAPE} with the policy gradient given by
\begin{align}
    \nabla J_1(\theta)&=\mathbb{E}_{\bm{\pi}}\left[\sum_i\nabla_{\theta_i}\log\pi_i(a_i|\tau_i)\mathbf{U}_i\right]\label{netdop_pg_before}\\ 
    &=\mathbb{E}_{\bm{\pi}}\left[\sum_{i,j}E_{ij}k_j(s)\nabla_{\theta_i}\log\pi_i(a_i|\tau_i)Q_j^{\phi_j}(s,a_j)\right], \label{netdop_pg}
\end{align}
where $k_j\geq0$ is the weight for agent $j$'s local $Q$ value $Q_j^{\phi_j}$ provided by the mixing network. The policy gradient derivation from Eq. \ref{netdop_pg_before} to Eq. \ref{netdop_pg} is provided in the appendix A. Since the local utility of other in-coalition agents is maximized by the policy updates, cooperation among agents is facilitated. Pseudo-code and more details of stochastic TAPE are provided in the appendix D.1.
\subsection{Deterministic TAPE}
Current deterministic MAPG methods \cite{peng2021facmac,zhang2021fop,zhou2022pac} yield fully-connected agent topology, which makes agents vulnerable to bad influence of other agents' sub-optimal actions. A mixing network $f_{\text{mix}}$ is adopted to mix local Q value functions $Q^{\pi_i}_i$. Each agent uses deterministic policy gradient to update parameters and directly maximize global $Q$ value $Q_{tot}^{\bm{\pi}}=f_{\text{mix}}(s,Q^{\pi_1}_1,\cdots,Q^{\pi_{n}}_{n})$. We use the agent topology to drop out utilities of out-of-coalition agents, so that influence of their sub-optimal actions will not spread to in-coalition agents. To this end, \emph{Coalition} $Q$ is defined as below.

\begin{mydef}[Coalition $Q$]
    Coalition $Q$ $Q^i_{\text{co}}$ for agent $i$ is the mixture of its in-coalition agents' local $Q$ values with mixing network $f_{\text{mix}}$, i.e.
\begin{equation}
    Q^i_{\text{co}}(s,\bm{a})=f_{\text{mix}}(s,\mathds{1}[E_{i1}]Q^{\pi_1}_1,\cdots,\mathds{1}[E_{i,n}]Q^{\pi_{n}}_{n}),
\end{equation}
where $\mathds{1}[E_{ij}]$ is the indicator function and $\mathds{1}[E_{ij}]=1$ only when edge $E_{ij}$ exists in the topology.
\end{mydef}\begin{figure*}[ht]
    \centering
    \includegraphics[width=.75\textwidth]{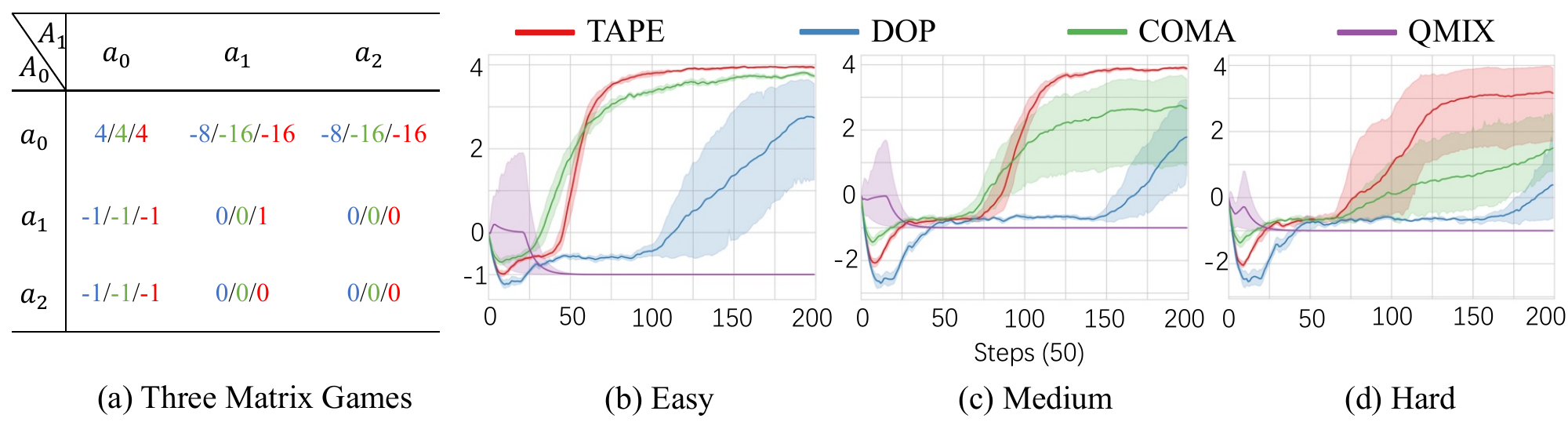}
    \caption{(a) gives the proposed three matrix games of different levels. We use different colors for different levels of game. Blue represents Easy, green represents Medium and red represents Hard. (b), (c) and (d) give evaluation results. Stochastic TAPE has the best performance because the agents directly maximize joint utility to achieve strong cooperation. The only difference between TAPE and DOP is that TAPE adopts the agent topology. Although COMA is seen as a weak baseline on SMAC, it achieves much better performance than DOP. QMIX fails to perform well in these games as they are not monotonic games.}
    \label{matrix_res}
\end{figure*}

During policy update, out-of-coalition agents' $Q$ values are always masked out, so agent $i$'s policy learning will not be affected by out-of-coalition agents. Based on Coalition $Q$, we propose \textbf{deterministic TAPE}, whose policy gradient is given by
\begin{align}
    \label{netpac_pg}
    \nabla J_2(\theta)&=\mathbb{E}_{\mathcal{D}}\left[\sum_i \nabla_{\theta_i}\pi_i(\tau_i)\nabla_{a_i}\hat{Q}_{\text{co}}^i(s,\bm{a})|_{a_i=\pi_i(\tau_i)}\right]
\end{align}
where $\hat{Q}_{\text{co}}^i(s,\bm{a})=f_{\text{mix}}\left(s,\mathds{1}[E_{i1}]\hat{Q}^{\phi_1}_1,\cdots,\mathds{1}[E_{i,n}]\hat{Q}^{\phi_{n}}_{n}\right)$ and $\hat{Q}^{\phi_i}_i(\tau_i,a_i,m_i)=Q^{\phi_i}_i(\tau_i,a_i,m_i)-\alpha\log\pi_i(a_i|\tau_i)$ is the local soft $Q$ value \cite{zhang2021fop} augmented with assistive information $m_i$ which contains information to assist policy learning towards the optimal policy as in \cite{zhou2022pac}.  After dropping out agents not in the coalition, the bad influence of out-of-coalition sub-optimal actions will not affect in-coalition agents. More details and pseudo-code are provided in the appendix D.2.
\section{Analysis}
\subsection{Agent Topology}
Although the agent topology can be any arbitrary topology, a proper agent topology should be able to explore diverse cooperation pattern, which is essential for robust cooperation \cite{li2021celebrating,strouse2021collaborating,lou2023pecan}. We studied three prevalent random graph model: Barabási–Albert (BA) model \cite{albert2002statistical}, Watts–Strogatz (WS) model \cite{watts1998collective} and Erdős–Rényi (ER) model \cite{erdHos1960evolution}. BA model is a scale-free network commonly used for citation and signaling biological networks \cite{barabasi1999emergence}. WS model is known as the small-world network where each nodes can be reached through a small number of nodes, resulting in the six degrees of separation \cite{travers1977experimental}. While in ER model, each edge between any two nodes has an independent probability of being present. Formally, the adjacency matrix $E$ of ER agent topology $(\mathcal{V},\mathcal{E})$ for $n$ agents is defined as $\forall i\in\{1,..,n\} $, $E_{ii}=1$; $\forall i,\ j\in\{1,..,n\}$, $i\neq j, E_{ij}=1$ with probability $p$ otherwise 0. 

In research question 1 of section \ref{smac_exp}, we found that ER model is able to generate the most diverse topologies, which in turn help the agents explore diverse cooperation pattern and achieve strongest performance. Thus, we use ER model to constitute the agent topology in the experiments.

\subsection{Theoretical Results}
We now establish policy improvement theorem of stochastic TAPE, and prove a theorem for the cooperation improvement from the perspective of exploring the parameter space, which is a common motivation in RL research \cite{schulman2017equivalence,haarnoja2018soft,zhang2021fop,adjodah2019leveraging}. We assume the policy to have tabular expressions.

The following theorem states that stochastic TAPE updates can monotonically improve the objective function $J(\bm{\pi})=\mathbb{E}_{\bm{\pi}}\left[\sum_t\gamma^tr_t\right]$.

\textbf{Theorem 1.} [stochastic TAPE policy improvement theorem] \emph{With tabular expressions for policies, for any pre-update policy $\bm{\pi}$ and updated policy $\hat{\bm{\pi}}$ by policy gradient in Eq. \ref{netdop_pg} that satisfy $\text{for any agent }i,\ \hat{\pi}_i(a_i|\tau_i)=\pi_i(a_i|\tau_i)+\beta_{a_i,s}\delta$, where $\delta$ is a sufficiently small number, we have
$J(\hat{\bm{\pi}})\geq J(\bm{\pi}),$
i.e. the joint policy is improved by the update.}

Please refer to Appendix B for the proof of Theorem 1. Although this policy improvement theorem is established for policies with tabular expressions, we provide conditions in the proof, under which policy improvement is guaranteed even with function approximators.

Next, we provide a theoretical insight that compared to using individual critics, stochastic TAPE can better explore the parameter space to find more effective cooperation pattern. One heuristic for measuring such capacity is the diversity of parameter updates during each iteration \cite{adjodah2019leveraging}, which is measured by the variance of parameter updates.\begin{figure*}[ht]
    \centering
    \includegraphics[width=.8\textwidth]{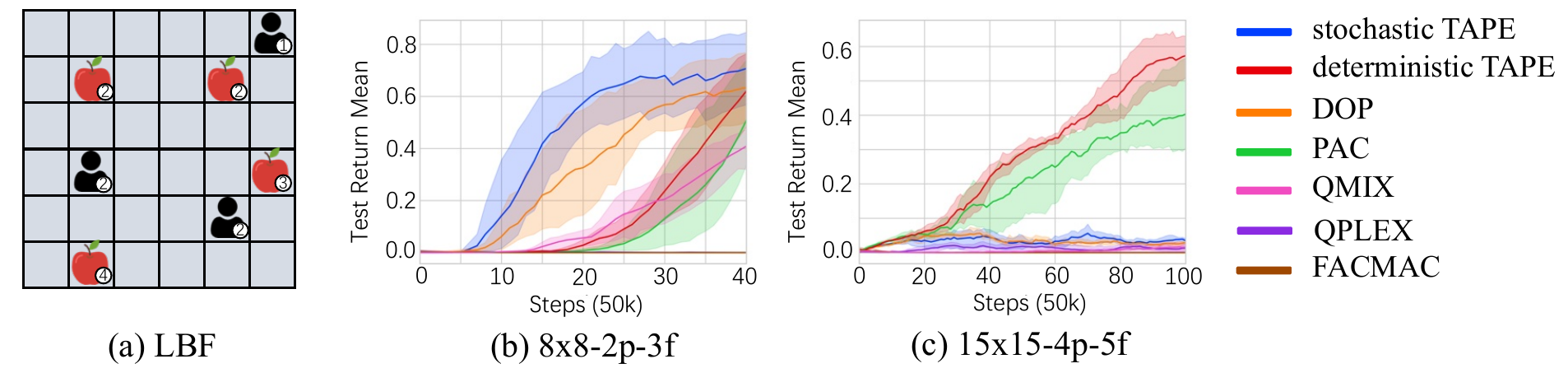}
    \caption{(a) gives a scenario 6x6-3p-4f in LBF. 6x6-3p-4f stands for 6x6 grid-world with 3 players and 4 fruits. (b) In 8x8-2p-3f, stochastic TAPE achieve best performance. While in the more difficult task 15x15-4p-5f (c), deterministic TAPE outperform its base method and all other baselines. See stochastic TAPE against DOP, and deterministic TAPE against PAC for comparison.}
    \label{lbf}
\end{figure*}

Given state $s$ and action $a_i$, let $\xi^\text{TAPE}_{a_i,s}$ and $\xi^\text{DOP}_{a_i,s}$ denote the stochastic TAPE and DOP parameter updates respectively. The following theorem states that stochastic TAPE policy update is more diverse so that it can explore the parameter space more effectively.

\textbf{Theorem 2.}  \emph{For any agent $i$ and $\forall s, a_i$, the stochastic TAPE policy update $\xi^\text{TAPE}_{a_i,s}$ and DOP policy update $\xi^\text{DOP}_{a_i,s}$ satisfy that $\text{Var}\left[\xi^\text{TAPE}_{a_i,s}\right]\geq\text{Var}\left[\xi^\text{DOP}_{a_i,s}\right]$, and $\Delta=\text{Var}\left[\xi^\text{TAPE}_{a_i,s}\right]-\text{Var}\left[\xi^\text{DOP}_{a_i,s}\right]$ is in proportion to $p^2$, where $p$ is the probability of edges being present in the Erdős–Rényi model, i.e. $\Delta\propto p^2.$}

Theorem 2 shows that compared to solely using individual critics, our agent topology provides larger diversity in policy updates to find better cooperation pattern. More details and proof are provided in the appendix C. It is worth noting that although a large hyperparameter $p$ in the agent topology means larger diversity in parameter updates, the CDM issue will also become severer because the connections among agents become denser. Thus, $p$ must be set properly to achieve compromise between facilitating cooperation and avoiding CDM issue, which we will show later in the experiments. 

\section{Experiment}
In this section, we first demonstrate that by ignoring other agents in the policy gradient to avoid bad influence of their sub-optimal actions, cooperation among agents is severely harmed. To this end, three one-step matrix games that require strong cooperation are proposed. Then, we evaluate the efficacy of the proposed methods on (a) Level-Based Foraging (LBF) \cite{papoudakis2021benchmarking}; (b) Starcraft II Multi-Agent Challenge (SMAC) \cite{samvelyan19smac}, and answer several research questions via various ablations and a heuristic graph search technique. Our code is available here\footnote{github.com/LxzGordon/TAPE}.
\subsection{Matrix Game}
We propose 3 one-step matrix games, which are harder versions of the example in introduction. The matrix games are given in Fig. \ref{matrix_res}(a). We use different colors to show rewards in different games (blue for Easy, green for Medium and red for Hard). The optimal joint policy is for both agents to take action $a_0$. But agent $A_0$ lacks motivation to choose $a_0$ because it is very likely to receive a large penalty ($-8$ or $-16$). Thus, this game requires strong cooperation among agents. In the Medium game, we further increase the penalty for agent 0 to choose $a_0$. In the Hard game, we keep the large penalty and add a local optimal reward at $(a_1,a_1)$. Note that these matrix games are not monotonic games \cite{rashid2020monotonic} as the optimal action for each agent depends on other agents. The evaluation results are given in Fig. \ref{matrix_res}.

With the agent topology to encourage cooperation, stochastic TAPE outperforms other methods by a large margin and is able to learn optimal joint policy even in the Hard game. DOP agents optimize individual utilities, ignoring utilities of other agents to avoid the influence of their sub-optimal actions, which result in severe miscoordination in these games. But since DOP agents adopt stochastic policy, they may receive some large reward after enough exploration. But the learning efficiency is much lower than stochastic TAPE. COMA is a weak baseline on complex tasks \cite{samvelyan19smac} (0\% win rate in all maps in section \ref{smac_exp}). But since COMA agents optimize global $Q$ value (expected team reward sum) instead of individual utility in DOP, it can achieve better results on these tasks requiring strong cooperation. These matrix games demonstrate the importance of considering the utility of other agents in cooperative tasks. With the agent topology, stochastic TAPE can facilitate cooperation among agents and alleviate CDM issue simultaneously.
\subsection{Level-Based Foraging}\begin{figure*}[ht]
    \centering
    \includegraphics[width=.83\textwidth]{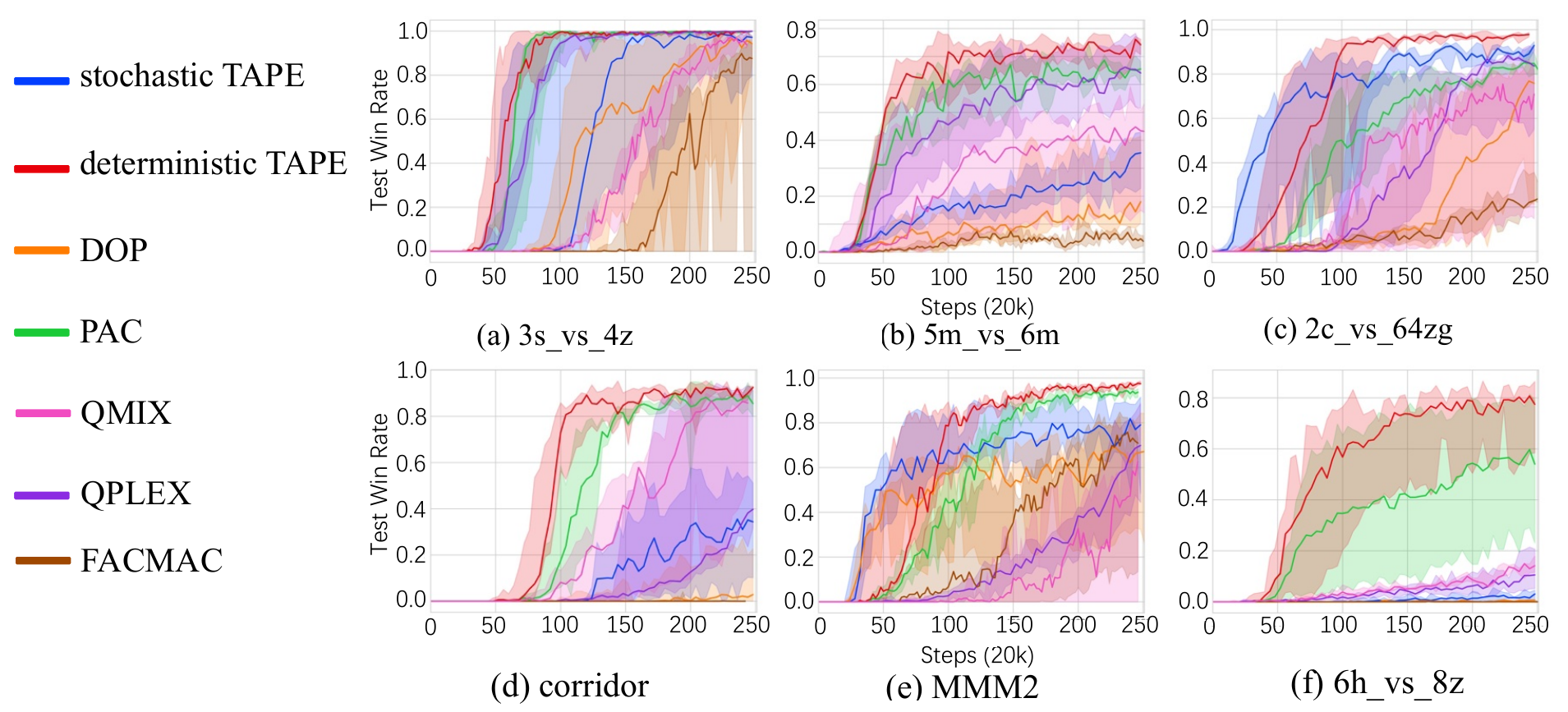}
    \caption{Experiment results on SMAC. (a-c) give the results in hard maps, and (d-f) are results in super-hard maps. After adopting our agent topology to facilitate cooperation and alleviate CDM issue, stochastic TAPE and deterministic TAPE outperforms their base methods respectively. See stochastic TAPE against DOP, and deterministic TAPE against PAC for comparison.}
    \label{smac_res}
\end{figure*}
In Level-Based Foraging (LBF \cite{papoudakis2021benchmarking}), agents navigate a grid-world and collect randomly-scattered food items. Agents and food items are assigned with levels. A food item is only allowed to be collected when near-by agents' level sum is larger than the food level. Reward is only given when a foot item is collected, assigning the environment with sparse-reward property. Test return is $1$ when all food items are collected. Compared baselines include both value-based methods: QMIX \cite{rashid2020monotonic} and QPLEX \cite{wang2020qplex}, and policy-based methods: DOP \cite{wang2020off}, FACMAC \cite{peng2021facmac} and PAC \cite{zhou2022pac}. Scenario illustration and results are given in Fig. \ref{lbf}.

To make 8x8-2p-3f more difficult, food items can only be collected when all agents participate. In this simple and sparse-reward task, with the stochastic policy and enhanced cooperation, stochastic TAPE outperforms all other methods on convergence speed and performance. While in 15x15-4p-5f, only state-of-the-art method PAC and deterministic TAPE learn to collect food items. With the agent topology to keep out bad influence of other agents' sub-optimal actions, deterministic TAPE achieves best performance.

\subsection{StarCraft Multi-Agent Challenge}\label{smac_exp}
StarCraft Multi-Agent Challenge (SMAC) \cite{samvelyan19smac} is a challenging benchmark built on StarCraft II, where agents must cooperate with each other to defeat enemy teams controlled by built-in AI. We evaluate the proposed methods and baselines with the recommended evaluation protocol and metric in six maps including three hard maps (3s\_vs\_4z, 5m\_vs\_6m and 2c\_vs\_64zg) and three super hard maps (corridor, MMM2 and 6h\_vs\_8z). All algorithms are run for four times with different random seeds. Each run lasts for $5\times10^6$ environmental steps. During training, each algorithm has four parallel environment to collect training data.

\begin{figure*}[t]
    \centering
    \includegraphics[width=.8\textwidth]{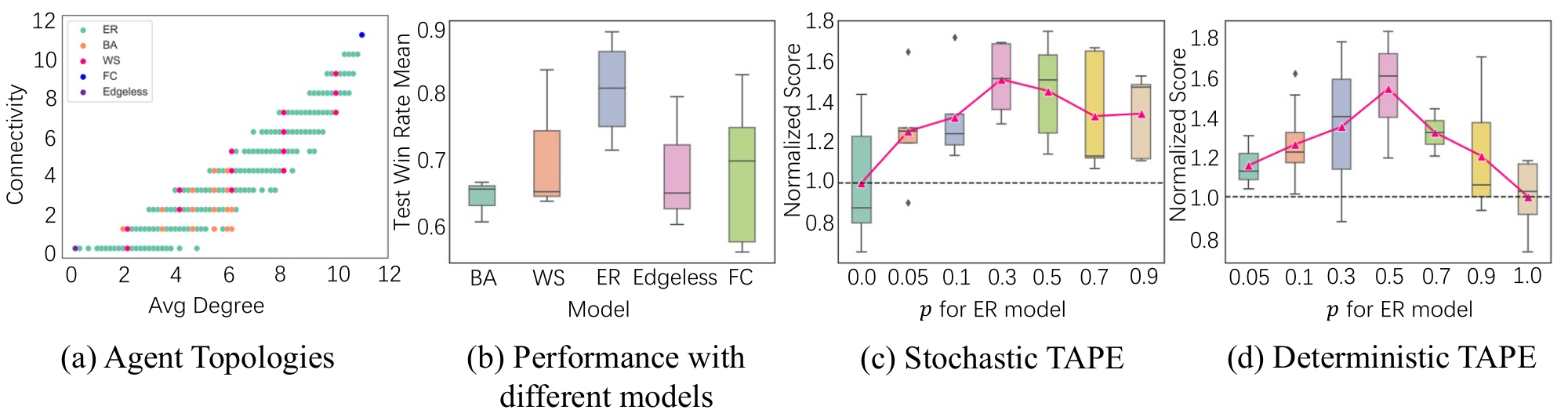}
    \caption{(a) and (b) show the results and performance of using different models to constitute agent topologies. {BA} is Barabási–Albert model, {WS} is Watts–Strogatz model, {ER} is Erdős–Rényi model, {Edgeless} and {FC} (Fully-Connected) are the topologies adopted in DOP and PAC respectively. ER has the most diverse topoloies and strongest performance. (c) and (d) show the performance of stochastic TAPE and deterministic TAPE in MMM2 with difference hyperparameter $p$ for ER model. Evaluation metric is test win rate and scores are normalized by the base method. In base method DOP, $p=0$ and base method PAC $p=1$. The boxplot is obtained with four different random seeds, and the red lines show the mean performance.}
    \label{abalation_other}
\end{figure*}\begin{figure*}[t]
    \centering
    \includegraphics[width=.73\textwidth]{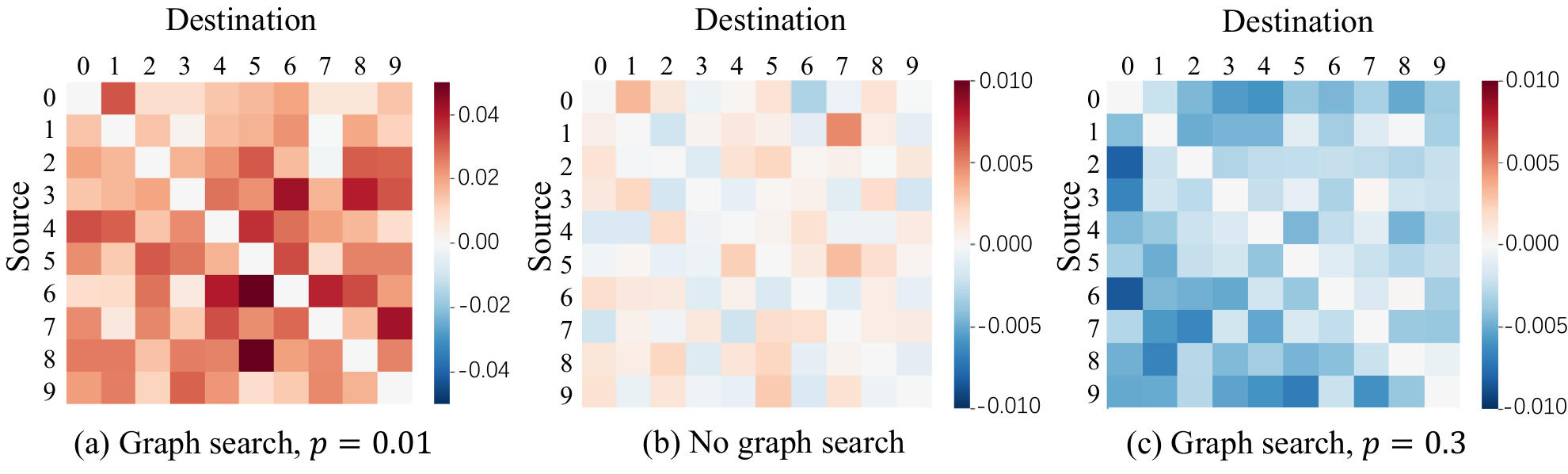}
    \caption{The heatmaps show the difference between the frequency of edges being present and the probability $p$. Source and Destination represent starting node and destination node of an edge. During training, over 1 million agent topology is generated. According to the law of large numbers, the difference is always around 0 when the heuristic graph search technique is not used in (b). In (a) and (c), we adopt the heuristic graph search technique to choose the agent topology with strongest performance. When $p$ is too small (0.01 in (a)), the connection among agents is too sparse, weakening cooperation among agents. Therefore, agent topologies with more edges can facilitate cooperation and are preferred by the graph search technique. As a results, the difference is always positive in (a). On the contrary, when the connection is too dense ($p=0.3$ in (c)), topologies with less edges are preferred because they stop bad influence of sub-optimal actions from spreading and have better performance, resulting in negative differences in (c).}
    \label{heatmap}
\end{figure*}

\textbf{Overall results} The overall results in six maps are provided in Fig. \ref{smac_res}. We can see deterministic TAPE outperforms all other methods in terms of performance and convergence speed. In 6h\_vs\_8z, one of the most difficult maps in SMAC, deterministic TAPE achieves noticeably better performance than its base method PAC and other baselines. It's worth noting that after integrating agent topology, both stochastic TAPE and deterministic TAPE have better performance compared to the base methods. This demonstrates the efficacy of the proposed agent topology in facilitating cooperation for DOP and alleviating CDM issue for PAC. Especially, in 2c\_vs\_64zg, stochastic TAPE outperforms all of the baselines except for our deterministic TAPE while its base method DOP struggles to perform well.

Next, we answer three research questions by ablations and additional experiments. The research questions are: \textbf{Q1.} What is the proper model to constitute the agent topology? \textbf{Q2.} Is there indeed a compromise between facilitating cooperation and suffering from the CDM issue? \textbf{Q3.} Is the agent topology capable of compromising between facilitating cooperation and the CDM issue to achieve best performance?

\textbf{Q1.} We study three prevalent random graph models: Barabási–Albert (BA) model \cite{albert2002statistical}, Watts–Strogatz (WS) model \cite{watts1998collective} and the Erdős–Rényi (ER) model \cite{erdHos1960evolution} via visualization and ablation study. First, we generate 1000 topologies for 12 agents with each model and give the visualization result in Fig. \ref{abalation_other}(a), where $x-$axis is average degree and $y-$axis is connectivity (minimum number of edges required, by removing which the graph becomes two sub-graphs). Average degree and connectivity are two essential factors for agent topology as they reflect the level of CDM issue and cooperation. Compared to the other two models, ER model generates much more diverse topologies, covering the area from edgeless topology to fully-connected topology. Then, we evaluate stochastic TAPE with each model on MMM2, a super hard map in SMAC. Results are given in Fig. \ref{abalation_other}(b). For the random graph models, the larger the graph diversity in Fig. \ref{abalation_other}(a), the stronger the performance is. Thus, we constitute the agent topology with ER model in other experiments. For fully-connected topology, the performance demonstrates very large variance, because once a sub-optimal action occurs, its bad influence will easily spread through the centralized critic to all other agents. It is worth nothing that the graphs can also be generated via Bayesian optimization, but this may also result in limited graph diversity, causing unstable or even worse performance. Thus, how to generate agent topology via optimization-based methods remains a challenge.

\textbf{Q2.} The compromise here means the more connection among agents to improve performance, the severer CDM issue becomes, and when it is too severe, it will in turn affect performance. To answer this research question, we devise a heuristic graph search technique. During policy training of agent $i$, we generate $n$ topologies with the ER model in each step and use them to update the agent policy. After obtaining $n$ updated policy $[\pi_1^i,..,\pi_n^i]$, we evaluate the post-update global $Q$ value $Q_{tot}^{\bm{\pi}^{-i},\pi^i_j}$ and choose the policy with largest global $Q$ value as the updated policy, i.e. $\pi^i=\arg\max_jQ_{tot}^{\bm{\pi}^{-i},\pi^i_j}$. The motivation of this heuristic graph search technique is that global $Q$ value is the expected future reward sum, which shows the post-update performance. Using this technique, we can find the topology with better performance. Then, we respectively use the graph search technique when $p$ is small or large and give the visualization of preferred topologies in Fig. \ref{heatmap}. The results confirm that the compromise does exist, because (1) facilitating cooperation by building more agent connections when there is little CDM issue (Fig. \ref{heatmap}(a)), and (2) removing connections to stop bad influence of sub-optimal actions from spreading when CDM issue is severe (Fig. \ref{heatmap}(c)), can both improve performance.

\textbf{Q3.} We answer this research question by giving the performance with different hyperparameter $p$, as it controls the level of CDM issue and cooperation. The results are given in Fig. \ref{abalation_other}(c), (d). Large $p$ stands for dense connections, where agents are easily affected by sub-optimal actions of other agents but cooperation is strongly encouraged. Small $p$ means sparse connections, where sub-optimal actions' influence will not easily spread but cooperation among agents is limited. (c) and (d) are drawn at the end of training and half of training to show the convergence performance and speed respectively. We can see the performances first increase when $p$ is small and later decrease when $p$ is too large. The best performance appears at the point where the cooperation is strong and CDM issue is acceptable. From the results, we can say our ER agent topology is able to compromise between cooperation and alleviating the CDM issue to achieve the best performance.

\section{Conclusion and Future Work}
In this paper, we propose an agent topology framework, which aims to alleviate the CDM issue without limiting agents' cooperation capacity. Based on the agent topology, we propose TAPE for both stochastic and deterministic MAPG methods. Theoretically, we prove the policy improvement theorem for stochastic TAPE and give a theoretical explanation about the improved cooperation among agents. Empirically, we evaluate the proposed methods on several benchmarks. Experiment results show that the methods outperform their base methods and other baselines in terms of convergence speed and performance. A heuristic graph search algorithm is devised and various studies are conducted, which validate the efficacy of our proposed agent topology.

\textbf{Limitation and Future Work} In this work, we consider constructing agent topology with existing random graph models without learning-based methods. Our future work is to adaptively learn the agent topology that can simultaneously facilitate agent cooperation and alleviate the CDM issue.


\bibliography{mybib}
\newpage
\section*{A$\quad$Derivation of Policy Gradient}
In this section, we give the derivation of the following policy gradient for stochastic TAPE
\begin{equation}
\begin{aligned}
    \nabla J(\theta)&=\mathbb{E}_{\bm{\pi}}\left[\sum_i\nabla_{\theta_i}\log\pi_i(a_i|\tau_i)\mathbf{U}_i\right]\\
    &=\mathbb{E}_{\bm{\pi}}\left[\sum_{i,j}E_{ij}k_j(s)\nabla_{\theta_i}\log\pi_i(a_i|\tau_i)Q_j^{\phi_j}(s,a_j)\right],
\end{aligned}
\end{equation}
where $\bm{U}_i=\sum\limits_{j=1}^nE_{ij}U_j$ is the coalition utility of agent $i$ with other agents connected in the agent topology, $U_j(s,a_j)=Q^\phi_{tot}(s,\bm{a})-\sum\limits_{a_j'}\pi_j({a_j'}|\tau_j)Q^\phi_{tot}(s,({a_j'},\bm{a}_{-j}))$ is the aristocrat utility of agent $j$ from \cite{wolpert2001optimal,wang2020off}, and $\forall i, E_{ii}=1$, i.e. all agents are connected with themselves.

\emph{Proof.}  First, we will reformulate the aristocrat utility for the policy gradient
\begin{equation*}
    \begin{aligned}
        &U_i(s,a_j)=Q^\phi_{tot}(s,\bm{a})-\sum\limits_{a_j'}\pi_i({a_j'}|\tau_i)Q^\phi_{tot}(s,({a_j'},\bm{a}_{-i})) \\
        &=\sum\limits_jk_j(s)Q^{\phi_j}_j(s,a_j)-\\&\sum\limits_{a_j'}\pi_i({a_j'}|\tau_i)\left[\sum\limits_{j\neq i}k_j(s)Q^{\phi_j}_j(s,a_j)+k_i(s)Q^{\phi_i}_i(s,{a_j'})\right]\\
        &=k_i(s)Q^{\phi_i}_i(s,a_i)-k_i(s)\sum\limits_{a_j'}\pi_i({a_j'}|\tau_i)Q^{\phi_i}_i(s,{a_j'})\\
        &=k_i(s)\left[Q^{\phi_i}_i(s,a_i)-\sum\limits_{a_j'}\pi_i({a_j'}|\tau_i)Q^{\phi_i}_i(s,{a_j'})\right].
    \end{aligned}
\end{equation*}
$U_j(s,a_j)=k_j(s)\left[Q_j^{\phi_j}(s,a_j)-\sum\limits_{a_j'}\pi_j({a_j'}|\tau_j)Q_j^{\phi_j}(s,{a_j'})\right]$. So policy gradient $g$ is

\begin{equation*}
\begin{aligned}
    g&= \mathbb{E}_{\bm{\pi}}\left[\sum_{i,j}E_{ij}\nabla_{\theta_i}\log\pi_i(a_i|\tau_i)U_j(s,a_j)\right]\\    &=\mathbb{E}_{\bm{\pi}}\Bigg[\sum_{i,j}E_{ij}\nabla_{\theta_i}\log\pi_i(a_i|\tau_i)k_j(s)\\
    &\left(Q_j^{\phi_j}(s,a_j)-\sum\limits_{a_j'}\pi_j({a_j'}|\tau_j)Q_j^{\phi_j}(s,{a_j'})\right)\Bigg].\\
\end{aligned}
\end{equation*}

Let $q_j(s)=\sum\limits_{a_j'}\pi_j({a_j'}|\tau_j)Q_j^{\phi_j}(s,{a_j'})$. Consider a given agent $i$
\begin{equation*}
    \begin{aligned}
        g_i&=\mathbb{E}_{\bm{\pi}}\left[\sum_jE_{ij}\nabla_{\theta_i}\log\pi_i(a_i|\tau_i)k_j(s)\left(Q_j^{\phi_j}(s,a_j)-q_j(s)\right)\right] \\        &=\mathbb{E}_{\bm{\pi}}\left[\sum_jE_{ij}\nabla_{\theta_i}\log\pi_i(a_i|\tau_i)k_j(s)Q_j^{\phi_j}(s,a_j)\right]-\\
        &\mathbb{E}_{\bm{\pi}}\left[\sum_jE_{ij}\nabla_{\theta_i}\log\pi_i(a_i|\tau_i)k_j(s)q_j(s)\right].\\
    \end{aligned}
\end{equation*}
Since
\begin{equation*}
    \begin{aligned}
        &\mathbb{E}_{\bm{\pi}}\left[\sum_jE_{ij}\nabla_{\theta_i}\log\pi_i(a_i|\tau_i)k_j(s)q_j(s)\right] \\
        =&\sum\limits_{s}d^{\bm{\pi}}(s)\sum\limits_{a_i}\sum\limits_{\bm{a}^{-i}}\pi_i(a_i|\tau_i)\bm{\pi}^{-i}(\bm{a}^{-i}|\bm{\tau}^{-i})\\
        &\sum_jE_{ij}\nabla_{\theta_i}\log\pi_i(a_i|\tau_i)k_j(s)q_j(s) \\
        =&\sum\limits_{s}d^{\bm{\pi}}(s)\sum\limits_{j}\sum\limits_{\bm{a}^{-i}}E_{ij}\bm{\pi}^{-i}(\bm{a}^{-i}|\bm{\tau}^{-i})k_j(s)q_j(s)\\
        &\sum\limits_{a_i}\pi_i(a_i|\tau)\nabla_{\theta_i}\log\pi_i(a_i|\tau_i)=0,
    \end{aligned}
\end{equation*}

$g_i=\mathbb{E}_{\bm{\pi}}\left[\sum_jE_{ij}\nabla_{\theta_i}\log\pi_i(a_i|\tau_i)k_j(s)Q_j^{\phi_j}(s,a_j)\right]$. Thus, 
\begin{equation*}
\nabla J(\theta)=\mathbb{E}_{\bm{\pi}}\left[\sum_{i,j}E_{ij}k_j(s)\nabla_{\theta_i}\log\pi_i(a_i|\tau_i)Q_j^{\phi_j}(s,a_j)\right]
\end{equation*}
is the policy gradient of stochastic TAPE.$\hfill\square$
\section*{B$\quad$Policy Improvement Theorem}
In this section, we give proof of the policy improvement theorem of stochastic TAPE. As in previous works \cite{degris2012off,wang2020off,feinberg2018model}, we relax the requirement that $Q^\phi_{tot}$ is a good estimate of $Q^{\bm{\pi}}_{tot}$ and simplify the $Q$-function learning process as the following MSE problem to make it tractable.
\begin{equation}
    L(\phi)=\sum\limits_{\bm{a},s}p(s)\bm{\pi}(\bm{a}|\bm{\tau})\left(Q^{\bm{\pi}}_{tot}(s,\bm{a})-Q^{\phi}_{tot}(s,\bm{a})\right)^2
\end{equation}
where $\bm{\pi}$ is the joint policy, $Q^{\bm{\pi}}_{tot}(s,\bm{a})$ is the true value and $Q^{\phi}_{tot}(s,\bm{a})$ is the estimated value.

To make this proof self-contained, we first borrow \textbf{Lemma 1} and \textbf{Lemma 2} from \cite{wang2020off}. \textbf{Lemma 1} states that the learning of centralized critic can preserve the order of local action values. Without loss of generality, we consider a given state $s$.

\textbf{Lemma 1.}  \emph{We consider the following optimization problem:}
\begin{equation}
    L_s(\phi)=\sum\limits_{\bm{a}}\bm{\pi}(\bm{a}|\bm{\tau})\left(Q^{\bm{\pi}}(s,\bm{a})-f(\bm{Q}^\phi(s,\bm{a}))\right)^2
\end{equation}
\emph{Here, $f(\bm{Q}^\phi(s,\bm{a})):R^n\rightarrow R$, and $\bm{Q}^\phi(s,\bm{a}))$ is a vector with the $i^{th}$ entry being $Q_i^{\phi_i}(s,a_i)$. $f$ satisfies that $\forall i, a_i, \frac{\partial f}{\partial Q^{\phi_i}_i(s,a_i)}>0$.}

\emph{Then, for any local optimal solution, it holds that}
\begin{equation*}
    Q^{\bm{\pi}}_i(s,a_i)\geq Q^{\bm{\pi}}_i(s,a'_i)\Longleftrightarrow Q^{\phi_i}_i(s,a_i)\geq Q^{\phi_i}_i(s,a'_i),\ \ \ \forall i,a_i,a'_i.
\end{equation*}

\emph{Proof.}  A  necessary condition for a local optimal is
\begin{equation*}
\begin{aligned}
        \frac{\partial L_s(\phi)}{\partial Q^{\phi_i}_i(s,a_i)}=\pi_i(a_i|\tau_i)\sum\limits_{\bm{a}_{-i}}\prod\limits_{j\neq i}\pi_j(a_j|\tau_j)\\
        \left(Q^{\bm{\pi}}(s,\bm{a})-f(\bm{Q}^\phi(s,\bm{a}))\right)(-\frac{\partial f}{\partial Q^{\phi_i}_i(s,a_i)})=0.
\end{aligned}
\end{equation*}
This implies that, for $\forall i, a_i$, we have
\begin{equation*}
    \begin{aligned}
        &\sum\limits_{\bm{a}_{-i}}\prod\limits_{j\neq i}\pi_j(a_j|\tau_j)\left(Q^{\bm{\pi}}(s,\bm{a})-f(\bm{Q}^\phi(s,\bm{a}))\right)=0\\
        \Rightarrow & \sum\limits_{\bm{a}_{-i}}\bm{\pi}_{-i}(\bm{a}_{-i}|\bm{\tau}_{-i})f\left(\bm{Q}^\phi\left(s,(a_i,\bm{a}_{-i})\right)\right)=Q_i^{\bm{\pi}}(s,a_i).
    \end{aligned}
\end{equation*}
Let $q(s,a_i)$ denote $\sum\limits_{\bm{a}_{-i}}\bm{\pi}_{-i}(\bm{a}_{-i}|\bm{\tau}_{-i})f\left(\bm{Q}^\phi\left(s,(a_i,\bm{a}_{-i})\right)\right)$. We have

\begin{equation*}
    \frac{\partial q(s,a_i)}{\partial Q_i^{\phi_i}(s,a_i)}=\sum\limits_{\bm{a}_{-i}}\bm{\pi}_{-i}(\bm{a}_{-i}|\bm{\tau}_{-i})\frac{f\left(\bm{Q}^\phi\left(s,(a_i,\bm{a}_{-i})\right)\right)}{\partial Q_i^{\phi_i}(s,a_i)} > 0.
\end{equation*}
Therefore, if $Q_i^{\bm{\pi}}(s,a_i)\geq Q_i^{\bm{\pi}}(s,a'_i)$, then any local minimal of $L_s(\phi)$ satisfies $Q^{\phi_i}_i(s,a_i)\geq Q^{\phi_i}_i(s,a'_i)$.$\hfill\square$

The mixer module of our method satisfies $\forall i, a_i, \frac{\partial f}{\partial Q^{\phi_i}_i(s,a_i)}>0$, so \textbf{Lemma 1} holds after the policy evaluation converges. 

\textbf{Lemma 2.}  \emph{For two sequences $\{a_i\}$, $\{b_i\}, i\in[n]$ listed in an increasing order. if $\sum_ib_i=0$, then $\sum_ia_ib_i\geq0$.}

\emph{Proof.}  We denote $\overline{a}=\frac{1}{n}\sum_ia_i$, then $\sum_ia_ib_i=\overline{a}(\sum_ib_i)+\sum_i\Tilde{a}_ib_i$, where $\sum_i\Tilde{a}_i=0$. Without loss of generality, we assume that $\overline{a}_i=0$, $\forall i$. $j$ and $k$ which $a_j\leq0,a_{j+1}\geq0$ and $b_k\leq0,b_{k+1}\geq0$. Since $a,b$ are symmetric, we assume $j\leq k$. Then, we have
\begin{equation*}
    \begin{aligned}
        \sum\limits_{i\in[n]}a_ib_i&=\sum\limits_{i\in[1,j]}a_ib_i+\sum\limits_{i\in[j+1,k]}a_ib_i+\sum\limits_{i\in[k+1,n]}a_ib_i \\
        &\geq \sum\limits_{i\in[j+1,k]}a_ib_i+\sum\limits_{i\in[k+1,n]}a_ib_i \\
        & \geq a_k\sum\limits_{i\in[j+1,k]}b_i+a_{k+1}\sum\limits_{i\in[k+1,n]}b_i.
    \end{aligned}
\end{equation*}
As $\sum\limits_{i\in[j+1,n]}b_i\geq0$, we have $-\sum_{i\in[j+1,k]}b_i\leq\sum_{i\in[k+1,n]}b_i$

Thus $\sum_{i\in[n]}a_ib_i\geq(a_{k+1}-a_{k})\sum_{i\in[k+1,n]}b_i\geq0$.$\hfill\square$

The next lemma states that for any policies with tabular expressions updated by stochastic TAPE policy gradient, the larger the local critic's value $Q_i^{\phi_i}(s,a_i)$, the larger update stepsize $\beta_{s,a_i}$ for $a_i$ under state $s$, and vice versa.

\textbf{Lemma 3.}  \emph{For any pre-update joint policy $\bm{\pi}$ and updated joint policy $\hat{\bm{\pi}}$ by stochastic TAPE policy gradient with tabular expressions that satisfy $\hat{\pi}_i(a_i|\tau_i)=\pi_i(a_i|\tau_i)+\beta_{a_i,s}\delta$ for any agent $i$, where $\delta$ is a sufficiently small number, $\forall s,a'_i,a_i$, it holds that}
\begin{equation}
\label{monotone_condition}
    Q_i^{\phi_i}(s,a_i)\geq Q_i^{\phi_i}(s,a'_i)\Longleftrightarrow \beta_{a_i,s}\geq\beta_{a'_i,s}.
\end{equation}

\emph{Proof.}  We start by showing the connection between $Q_i^{\phi_i}(s,a_i)$ and $\beta_{a_i,s}$. $\nabla_{\theta_i}J(\theta)$ is the policy gradient for agent $i$
\begin{equation*}
    \begin{aligned}
        \nabla_{\theta_i}J(\theta)&=\mathbb{E}_{\bm{\pi}}\left[\sum_jE_{ij}\nabla_{\theta_i}\log\pi_i(a_i|\tau_i)k_j(s)Q_j^{\phi_j}(s,a_j)\right] \\
        &=\sum\limits_{s}d^{\bm{\pi}}(s)\sum\limits_{a_i}\sum\limits_{\bm{a}^{-i}}\pi_i(a_i|\tau_i)\bm{\pi}^{-i}(\bm{a}^{-i}|\bm{\tau}^{-i})\\
        &\qquad\qquad\sum\limits_jE_{ij}\nabla_{\theta_i}\log\pi_i(a_i|\tau_i)k_j(s)Q_j^{\phi_j}(s,a_j) \\
        &=\sum\limits_{s,\bm{a}^{-i}}d^{\bm{\pi}}(s)\bm{\pi}^{-i}(\bm{a}^{-i}|\bm{\tau}^{-i})\\
        &\qquad\qquad\sum\limits_{a_i}\sum\limits_jE_{ij}\nabla_{\theta_i}\pi_i(a_i|\tau_i)k_j(s)Q_j^{\phi_j}(s,a_j).
    \end{aligned}
\end{equation*}

With a little abuse of notation, we let $d(s,\bm{\tau}^{-i})=d^{\bm{\pi}}(s)\bm{\pi}^{-i}(\bm{a}^{-i}|\bm{\tau}^{-i})$ for simplicity, where $\bm{\tau}$ is joint agent action-observation history and $-i$ stands for excluding agent $i$. Thus, without loss of generality, consider some given state $s$ and joint action $\bm{a}$, the policy gradient $g_i$ for agent $i$ is
\begin{equation}
\label{g_i}
    \begin{aligned}
        g_i&=d(s,\bm{\tau}^{-i})\sum\limits_jE_{ij}\nabla_{\theta_i}\pi_i(a_i|\tau_i)k_j(s)Q_j^{\phi_j}(s,a_j) \\
        &=d(s,\bm{\tau}^{-i})\nabla_{\theta_i}\pi_i(a_i|\tau_i)k_i(s)Q_i^{\phi_i}(s,a_i)+\\
        &\qquad\qquad d(s,\bm{\tau}^{-i})\sum\limits_{j\neq i}E_{ij}\nabla_{\theta_i}\pi_i(a_i|\tau_i)k_j(s)Q_j^{\phi_j}(s,a_j). \\
    \end{aligned}
\end{equation}

Since the policies are with tabular expressions, then $\pi_i(a_i|\tau_i)=\theta_{a_i,\tau_i}$. And the update $\beta_{a_i,s}$ is in proportion to the gradient, so that
$$\beta_{a_i,s}\propto g_i=d(s,\bm{\tau}^{-i})k_i(s)Q_i^{\phi_i}(s,a_i)+C.$$
where $C$ is a constant independent of $a_i$ and $Q_i^{\phi_i}(s,a_i)$.

Since $k_i\geq0$ is a positive coefficient given by the mixing network, $d(s,\bm{\tau}^{-i})k_i(s)Q_i^{\phi_i}(s,a_i)+C$ is a linear function with positive coefficient w.r.t. $Q_i^{\phi_i}(s,a_i)$. Thus $\forall s,a'_i,a_i$, it holds that $Q_i^{\phi_i}(s,a_i)\geq Q_i^{\phi_i}(s,a'_i)\Longleftrightarrow \beta_{a_i,s}\geq\beta_{a'_i,s}.\hfill\square$

Now, we are ready to provide proof for the policy improvement theorem.

\textbf{Theorem 1.}  \emph{With tabular expressions for policies, for any pre-update policy $\bm{\pi}$ and updated policy $\hat{\bm{\pi}}$ by policy gradient of stochastic TAPE that satisfy $\text{for any agent }i,\ \hat{\pi}_i(a_i|\tau_i)=\pi_i(a_i|\tau_i)+\beta_{a_i,s}\delta$, where $\delta$ is a sufficiently small number, we have
$J(\hat{\bm{\pi}})\geq J(\bm{\pi}),$
i.e. the joint policy is improved by the update.}

\emph{Proof.}  Given a good value estimate $Q^\phi_{tot}$, from \textbf{Lemma 1} and \textbf{Lemma 3} we have 
\begin{equation}
\label{monotocity}
    Q^{\bm{\pi}}_i(s,a_i)\geq Q^{\bm{\pi}}_i(s,a'_i) \Longleftrightarrow \beta_{a_i,s}\geq\beta_{a'_i,s}.
\end{equation}
Given $\forall s_t$, we have
\begin{equation*}
    \begin{aligned}
        &\qquad\sum\limits_{\bm{a}_t}\hat{\bm{\pi}}(\bm{a}_t|\bm{\tau}_t)Q^{\bm{\pi}}_{tot}(s_t,\bm{a}_t)\\
        &=\sum\limits_{\bm{a}_t}\left(\prod\limits^n_i\hat{\pi}_i(a_i^t|\tau_i^t)\right)Q^{\bm{\pi}}_{tot}(s_t,\bm{a}_t) \\
        &=\sum\limits_{\bm{a}_t}\left(\prod\limits^n_i{\pi_i(a_i^t|\tau_i^t)+\beta_{a_i^t,s_t}\delta}\right)Q^{\bm{\pi}}_{tot}(s_t,\bm{a}_t) \\
        &=\sum\limits_{\bm{a}_t}\prod\limits^n_i\pi_i(a_i^t|\tau_i^t)Q^{\bm{\pi}}_{tot}(s_t,\bm{a}_t)+\\
        &\qquad\qquad\delta\sum\limits^n_{i=1}\sum\limits_{a_i^t}\beta_{a_i^t,s_t}Q^{\bm{\pi}}_{tot}(s_t,a_i^t)+o(\delta)\\
        &=V^{\bm{\pi}}_{tot}(s_t)+\delta\sum\limits^n_{i=1}\sum\limits_{a_i^t}\beta_{a_i^t,s_t}Q^{\bm{\pi}}_{tot}(s_t,a_i^t)+o(\delta).
    \end{aligned}
\end{equation*}
Since $\delta$ is sufficiently small, we use $o(\delta)$ to represent the summation of components where the exponential coefficient of $\delta$ is greater than $1$. $o(\delta)$ is omitted in further analysis since it is sufficiently small.

Because $\sum\limits_{a_i^t}\left[\pi_i(a_i^t|s_t)+\beta_{a_i^t,s_t}\right]=1$, we have $\sum\limits_{a_i^t}\beta_{a_i^t,s_t}=0$. From \textbf{Lemma 2} and Eq. \ref{monotocity}, we have $\sum\limits^n_{i=1}\sum\limits_{a_i^t}\beta_{a_i^t,s_t}Q^{\bm{\pi}}_{tot}(s_t,a_i^t)>0$. Thus 
\begin{equation*}
    \sum\limits_{\bm{a}_t}\hat{\bm{\pi}}(\bm{a}_t|\bm{\tau}_t)Q^{\bm{\pi}}_{tot}(s_t,\bm{a}_t)\geq V^{\bm{\pi}}_{tot}(s_t).
\end{equation*}
The rest of the proof follows the policy improvement theorem for tabular MDPs from \cite{sutton2018reinforcement}.
\begin{equation*}
    \begin{aligned}
        V^{\bm{\pi}}_{tot}(s_t) &\leq \sum\limits_{\bm{a}_t}\hat{\bm{\pi}}(\bm{a}_t|\bm{\tau}_t)Q^{\bm{\pi}}_{tot}(s_t,\bm{a}_t) \\
        &=\sum\limits_{\bm{a}_t}\hat{\bm{\pi}}(\bm{a}_t|\bm{\tau}_t)\left(r_t+\gamma\sum\limits_{s_{t+1}}p(s_{t+1}|s_t,\bm{a}_t)V^{\bm{\pi}}(s_{t+1})\right) \\
        & \leq \sum\limits_{\bm{a}_t}\hat{\bm{\pi}}(\bm{a}_t|\bm{\tau}_t)\Bigg(r_t+\gamma\sum\limits_{s_{t+1}}p(s_{t+1}|s_t,\bm{a}_t)\\
        &\qquad\qquad\Big(\sum\limits_{\bm{a}_{t+1}}\bm{\pi}(\bm{a}_{t+1}|\bm{\tau}_{t+1})Q^{\bm{\pi}}_{tot}(s_{t+1},\bm{a}_{t+1})\Big)\Bigg)\\
        &\cdots \\
        &= V^{\hat{\bm{\pi}}}_{tot}(s_t).
    \end{aligned}
\end{equation*}
So we have 
$$V^{\hat{\bm{\pi}}}_{tot}(s_0) \geq V^{\bm{\pi}}_{tot}(s_0), \ \ \ \forall s_0\\$$   
   $$\Longrightarrow \ \ \ \sum\limits_{s_0}p(s_0)V^{\hat{\bm{\pi}}}_{tot}(s_0) \geq \sum\limits_{s_0}p(s_0)V^{\bm{\pi}}_{tot}(s_0),$$

Which is equivalent to $J(\hat{\bm{\pi}})\geq J(\bm{\pi})$, since $J(\bm{\pi})=\sum\limits_{s_0}p(s_0)V^{\bm{\pi}}_{tot}(s_0). \hfill\square$

\textbf{Remark} We prove that with the coalition utility $\bm{U}_i$ in the policy gradient, the objective function $J(\bm{\pi})$ is monotonically maximized. The monotone condition (Eq. \ref{monotone_condition}) guarantees the monotonic improvement of stochastic TAPE policy updates in tabular cases. In cases where function approximators (such as neural networks) are used, the policy improvements are still guaranteed as long as the monotone condition holds (actions with larger values have larger update stepsizes). In the experiment section, we empirically demonstrate that the policies parameterized by deep neural networks have steady performance improvement as training goes on, and agents have better performance compared to the baselines.

\section*{C$\quad$Policy Update Diversity}
In this section, we give the detailed proof of \textbf{Theorem 2}. We define the diversity of exploration in the parameter space as the variance of parameter updates $\xi_{a_i,s}^\text{TAPE}=\mathbb{E}_E\left[g_{a_i,s}^{\text{TAPE}}\lambda\right],\ \xi_{a_i,s}^\text{DOP}=g_{a_i,s}^{\text{DOP}}\lambda$, where $g_{a_i,s}$ is the policy gradient given $s$ and $a_i$, $\lambda$ is learning rate and $E$ is the Erdős–Rényi network in stochastic TAPE. Without loss of generality, we assume $\lambda=1$.

For clarity, we first restate \textbf{Theorem 2}.

\textbf{Theorem 2.}  \emph{For any agent $i$ and $\forall s, a_i$, the stochastic TAPE policy update $\xi^\text{TAPE}_{a_i,s}$ and DOP policy update $\xi^\text{DOP}_{a_i,s}$ satisfy that $\text{Var}\left[\xi^\text{TAPE}_{a_i,s}\right]\geq\text{Var}\left[\xi^\text{DOP}_{a_i,s}\right]$, and $\Delta=\text{Var}\left[\xi^\text{TAPE}_{a_i,s}\right]-\text{Var}\left[\xi^\text{DOP}_{a_i,s}\right]$ is in proportion to $p^2$, where $p$ is the probability of edges being present in the Erdős–Rényi model, i.e. $\Delta\propto p^2.$}

\emph{Proof.}  From the proof of \textbf{Lemma 3}, we have
$$g_{a_i,s}^{\text{TAPE}}=d(s,\bm{\tau}^{-i})\sum\limits_jE_{ij}k_j(s)Q_j^{\phi_j}(s,a_j).$$
By replacing the adjacency matrix $E$ with identity matrix $I$, we have 
$$g_{a_i,s}^{\text{DOP}}=d(s,\bm{\tau}^{-i})k_i(s)Q_i^{\phi_i}(s,a_i).$$
Substitute $g_{a_i,s}^{\text{TAPE}}$ into $\xi^\text{TAPE}_{a_i,s}$
\begin{equation*}
\begin{aligned}
        \xi^\text{TAPE}_{a_i,s}&=\mathbb{E}_E\Big[d(s,\bm{\tau}^{-i})\nabla_{\theta_i}\pi_i(a_i|\tau_i)k_i(s)Q_i^{\phi_i}(s,a_i)+\\
        &\qquad\quad\sum\limits_{j\neq i}d(s,\bm{\tau}^{-i})E_{ij}\nabla_{\theta_i}\pi_i(a_i|\tau_i)k_j(s)Q_j^{\phi_j}(s,a_j)\Big]\\
        &=d(s,\bm{\tau}^{-i})k_i(s)Q_i^{\phi_i}(s,a_i)+\\
        &\qquad\qquad\mathbb{E}_E\left[\sum\limits_{j\neq i}d(s,\bm{\tau}^{-i})E_{ij}k_j(s)Q_j^{\phi_j}(s,a_j)\right] \\
        &=\beta_{a_i,s}^{\text{DOP}}+\sum\limits_{j\neq i}\mathbb{E}_E\left[d(s,\bm{\tau}^{-i})E_{ij}k_j(s)Q_j^{\phi_j}(s,a_j)\right]\\
        &=\xi_{a_i,s}^{\text{DOP}}+p\sum\limits_{j\neq i}d(s,\bm{\tau}^{-i})k_j(s)Q_j^{\phi_j}(s,a_j),\\
\end{aligned}
\end{equation*}
where $d(s,\bm{\tau}^{-i})k_j(s)Q_j^{\phi_j}(s,a_j)=C_j$ is independent of $\pi_i$ and $Q_i^{\phi_i}$. Thus,
\begin{equation*}
    \begin{aligned}
        \text{Var}\left[\xi^\text{TAPE}_{a_i,s}\right]&=\text{Var}\left[\xi_{a_i,s}^{\text{DOP}}+p\sum\limits_{j\neq i}C_j\right]\\
        &=\text{Var}\left[\xi_{a_i,s}^{\text{DOP}}\right]+p^2\sum\limits_{j\neq i}\text{Var}\left[C_j\right].
    \end{aligned}
\end{equation*}
Since variance is always non-negative, clearly we have $\text{Var}\left[\xi^\text{TAPE}_{a_i,s}\right]\geq\text{Var}\left[\xi^\text{DOP}_{a_i,s}\right]$
and $\Delta\propto p^2$.$\hfill\square$

\textbf{Remark}  The above theorem states that stochastic TAPE explores the parameter space more effectively. This provides a theoretical insight and explanation why stochastic TAPE agents are more capable of finding good cooperation patterns with other agents. And $\Delta$ is in proportion to $p^2$, which means as $p$ increase, the connections among agents in the topology become denser and enhance their capability of cooperation. But the dense connection will also introduces the CDM issue, as sub-optimality of some agents will more easily affect other agents. Thus, $p$ serves as a hyperparameter to compromise between avoiding CDM issue and capability to explore cooperation patterns.

\section*{D$\quad$Algorithm}
In this section, we give pseudo-code of the proposed methods. As we only modify policy gradients, rest of structures remains the same as the base methods \cite{wang2020off,zhou2022pac}.
\subsection*{D.1$\quad$Stochastic TAPE}
We first give the pseudo-code of stochastic TAPE and then give the details.
\begin{algorithm}[h]
        \caption{Stochastic TAPE}
        \begin{algorithmic}[1]
            \State Initialize critic $\phi$, target critic $\phi'=\phi$, policy $\theta_i$ for each agent $i$, and off-policy replay buffer $\mathcal{D}$
            \While{training not finished}
                \State Rollout $n$ trajectories and store them in $\mathcal{D}$
                \State Use the trajectories to calculate $y^\text{on}$
                \State Sample a batch of trajectories from $\mathcal{D}$ to calculate $y^\text{off}$
                \State Update $\phi$ with $y^\text{on}$ and $y^\text{off}$ according to Eq. \ref{dop_critic_loss}
                \State Generate an agent topology $E$ according to Eq. \ref{supp_erdos}
                \State Update policy parameter $\theta$ with the on-policy trajectories according to Eq. \ref{tape_stochastic_loss}
                \State Copy critic network parameter $\phi$ to target critic $\phi'$ every $m$ episode
            \EndWhile
        \end{algorithmic}
\end{algorithm}

The policy gradient of stochastic TAPE is given by
\begin{equation}\label{tape_stochastic_loss}
    g=\mathbb{E}_{\bm{\pi}}\left[\sum_{i,j}E_{ij}k_j(s)\nabla_{\theta_i}\log\pi_i(a_i|\tau_i)Q_j^{\phi_j}(s,a_j)\right].
\end{equation}
where $E$ is the ER agent model, defined as 
\begin{equation}
\label{supp_erdos}
    \begin{aligned}
    &\ \ \ \ \ \ \ \forall\ i,j\in\{1,..,n\}\\ 
    &\text{if}\ i= j,\ \ \ \ E_{ij}=1 \\ 
    &\text{else}\ E_{ij}=\left\{
        \begin{aligned}
            &1\ \ \ \rm{with\ probability}\ \emph{p} \\
            &0\ \ \ \rm{otherwise}
        \end{aligned}
            \right..
    \end{aligned}
\end{equation}

And as in the base method DOP, stochastic TAPE adopts an off-policy critic to improve sample efficiency, where the critics' training loss is the mixture of an off-policy loss with target $y^\text{off}$ based on tree-backup technique \cite{precup2000eligibility,munos2016safe} and an on-policy loss with target $y^\text{on}$, i.e.
\begin{align}
    \label{dop_critic_loss}
    &\mathcal{L}(\phi)=\kappa\mathbb{E}_\mathcal{D}\left[\text{MSE}(y^\text{off},Q^\phi_{tot})\right]+(1-\kappa)\mathbb{E}_{\bm{\pi}}\left[\text{MSE}(y^\text{on},Q^\phi_{tot})\right],
\end{align}
where
\begin{equation}
\label{dop_off_target}
\begin{aligned}
    &y^\text{off}=Q^{\phi'}_{tot}(s,\bm{a})+\sum\limits_{t=0}^{k-1}\gamma^tc_t\Bigg[r_t+\\
    &\qquad\gamma\sum_ik_i(s_{t+1})\mathbb E_{\pi_i}\left[Q_i^{\phi_i'}(s_{t+1},\cdot)\right]+b(s_{t+1})-Q^{\phi'}_{tot}(s_t,\bm{a}_t)\Bigg]
\end{aligned}
\end{equation}
\begin{equation}
    \begin{aligned}
    &y^\text{on}=Q^{\phi'}_{tot}(s,\bm{a})+\sum\limits_{t=0}^\infty(\gamma\lambda)^t\Bigg[r_t+\gamma Q^{\phi'}_{tot}(s_{t+1},\bm{a}_{t+1})-Q^{\phi'}_{tot}(s_t,\bm{a}_t)\Bigg]. \label{dop_on_target}
\end{aligned}
\end{equation}

where MSE is the mean-squared error loss function, $\kappa$ is a parameter controlling the importance of off-policy learning, $\mathcal{D}$ is the off-policy replay buffer, $\bm{\pi}$ is the joint policy, $Q^{\phi}_{tot}(s,\bm{a})=\sum_ik_i(s_t)Q_i^{\phi_i}(s_{t},a^i_t)+b(s_{t})$, $k\geq0$ and $b$ are coefficients provided by the mixing network, $Q^{\phi'}$ is the target network to stabilize training \cite{mnih2013playing}, $c_t=\prod_{l=1}^t\lambda\bm{\pi}(\bm{a}_l|s_l)$ and $\lambda$ is the TD($\lambda$) hyperparameter.

With all the equations above, the pseudo-code of stochastic TAPE is given in Algorithm 1.
\subsection*{D.2$\quad$Deterministic TAPE}
We first give the pseudo-code of deterministic TAPE and then give the details.
\begin{algorithm}[h]
        \caption{Deterministic TAPE}
        \begin{algorithmic}[1]\label{det}
            \State Initialize critic $\phi_i$, target critic $\phi'_i=\phi_i$, policy $\theta_i$ for each agent $i$, mixing network $f_{\text{mix}}$, information encoder $f_m$ and replay buffer $\mathcal{D}$
            \While{training not finished}
                \State Rollout $n$ trajectories and store them in $\mathcal{D}$
                \State Sample a batch of trajectories from $\mathcal{D}$
                \State Sample assistive information $m_i\sim N(f_m(o_i),I)$ for all agents
                \State Calculate $\mathcal{L}_{\bm{\pi}}$, $\mathcal{L}_{CA}$, $\mathcal{L}_{IB}$, $\mathcal{L}_{\hat{Q}^*}$ and $\mathcal{L}_{Q_{tot}}$ according to Eq. \ref{pi_loss}, \ref{q*_loss}, \ref{q_tot_loss}, \ref{ca_loss}, \ref{ib_loss}
                \State Total loss $\mathcal{L}=-\mathcal{L}_{\bm{\pi}}+\mathcal{L}_{CA}+\mathcal{L}_{IB}+\mathcal{L}_{\hat{Q}^*}+\mathcal{L}_{Q_{tot}}$
                \State Update the critics, mixing network, policy networks and information encoder to minimize $\mathcal{L}$
                \State Copy critic network parameter $\phi$ to target critic $\phi'$ every $m$ episode
            \EndWhile
        \end{algorithmic}
\end{algorithm}

The policy gradient of deterministic TAPE is given by
\begin{equation}
    \label{pi_loss}
    \begin{aligned}
        &\mathcal{L}_{\bm{\pi}}=\mathbb{E}_{\mathcal{D}}\Bigg[\sum_i \nabla_{\theta_i}\pi_i(\tau_i)\\
        &\qquad\nabla_{a_i}f_{\text{mix}}\left(s,\mathbbm{1}[E_{i1}]\hat{Q}^{\phi_1}_1,\cdots,\mathbbm{1}[E_{i,n}]\hat{Q}^{\phi_{n}}_{n}\right)|_{a_i=\pi_i(\tau_i)}\Bigg].
    \end{aligned}
\end{equation}
where $\hat{Q}^{\phi_i}_i(\tau_i,a_i,m_i)=Q^{\phi_i}_i(\tau_i,a_i,m_i)-\alpha\log\pi_i(a_i|\tau_i)$ is the soft $Q$ network augmented with assistive information $m_i$. The assistive information $m_i$ is encoded from the observation $o_i$ of agent $i$, which provides information about the optimal joint action $\bm{a}^*$ and assists the value factorization.

PAC follows the design of WQMIX \cite{rashid2020weighted} and keep two mixing network $Q_{tot}$ and $\hat{Q}^*$. $Q_{tot}$ is the monotonic mixing network as in QMIX \cite{rashid2020monotonic}, while $\hat{Q}^*$ is an unrestricted function to make sure the joint-action values are correctly estimated as in WQMIX. The loss functions for $Q_{tot}$ and $\hat{Q}^*$ are given by
\begin{align}
    \mathcal{L}_{\hat{Q}^*}&=\sum_{k=0}^N(\hat{Q}^*(s_k,\hat{\bm{a}}_k)-y_k)^2\label{q*_loss} \\
    \mathcal{L}_{Q_{tot}}&=\sum_{k=0}^N\omega(s_k,\bm{a}_k)(Q_{tot}(s_k,\bm{a}_k,\bm{m}_k)-y_k)^2\label{q_tot_loss}.
\end{align}
where $N$ is the batch size, $\hat{a}_i=\arg\max Q_i^{\phi_i}(\tau_i,\cdot,m_i)$, $\hat{\bm{a}}=[\hat{a}_1,..,\hat{a}_n]$, $\omega(s,\bm{a})$ is the weighting function in WQMIX, $\bm{m}$ is the joint assistive information, $Q_{tot}$ is the mixture of local $Q$ values and target $y_k=r_k+\gamma\hat{Q}^*(s'_k,\arg\max_{\hat{\bm{a}}'_k} Q_{tot}(s'_k,\hat{\bm{a}}'_k,\bm{m}'_k;\phi'))$ with $\phi'$ being the parameters of the target network. PAC adopts two auxiliary loss to assist value factorization in MAPG, which we will briefly introduce next.

\textbf{Counterfactual Assistance Loss:} The counterfactual assistance loss is proposed to directly guide individual agent's policy towards the action $\hat{a}_i^*$ from $\hat{Q}^*$. To this end, they propose an advantage function with a counterfactual baseline that relegates $\hat{a}_i^*$ while keeping other all other agents' actions $\bm{a}^{-i}$ fixed. Thus, the counterfactual assistance loss is given by
\begin{equation}\label{ca_loss}
\begin{aligned}
        &\mathcal{L}_{CA}=\sum_i\log\pi_i(a_i|\tau_i)\\
    &\qquad\sum_{a_i}\left[Q^{\phi_i}_i(\tau_i,\hat{a}^*_i,m_i)-\pi_i(a_i|\tau_i)Q^{\phi_i}_i(\tau_i,a_i,m_i)\right].
\end{aligned}
\end{equation}
 
\textbf{Information Bottleneck Loss:} Inspired by information bottleneck method \cite{tishby2000information}, the information bottleneck loss encodes the optimal joint action $\hat{\bm{a}}^*$ as the assistive information $m_i$ for local $Q$ value functions $Q_i^{\pi_i}(\tau_i,a_i,m_i)$. The assistive information is maximally informative about the optimal action $a_i^*$. With the deep variational information bottleneck \cite{alemi2016deep}, the variational lower bound of this objective is
\begin{equation}\label{ib_loss}
\begin{aligned}
    &\mathcal{L}_{IB}=\mathbb{E}_{o_i\sim\mathcal{D},m_j\sim f_m}\Bigg[-\mathcal{H}\left[p(\hat{a}^*_j|\bm{o}),q_\psi(\hat{a}^*_j|o_j,\bm{m})\right]+\\
    &\qquad\qquad\qquad\qquad\beta D_{KL}(p(m_i|o_i)\|q_\sigma(m_i)\Bigg],
\end{aligned}
\end{equation}
where $\mathcal{H}$ is the entropy operator, $D_{KL}$ is the KL divergence and $q_\sigma(m_i)$ is a variational posterior estimator of $p(m_i)$ with parameter $\sigma$. The information encoder $f_m$ is trained to encode assistive information $m_i\sim N(f_m(o_i;\theta_m),I)$, where $N$ is the normal distribution.

With all the loss terms defined above, pseudo-code of deterministic TAPE is given in Algorithm 2. 
\section*{E$\quad$Experiment and Implementation Details}
We run experiments on Nvidia RTX Titan graphics cards with an Intel Xeon Gold 6240R CPU. The curves in our experiments are smoothed by a sliding window, and the window size is 4, i.e. results of each timestep is the average of the past 4 timesteps. More experiment and implementation details in each environment are given below. It is worth noting that the only hyperparameter that needs tuning in our methods is $p$, the probability of edges being present.

\subsection*{E.1$\quad$Matrix Game} 
The evaluation metric in the matrix game is the average return of last 100 episodes and the training goes on for 10000 episodes in total. The results are drawn with four random seeds, which are randomly initialized at the beginning of an experiment. For this simple matrix game, we use tabular expressions for policies in stochastic TAPE, DOP and COMA. The critics are parameterized by a three-layer feed-forward neural network with hidden size 32. The mixing network for QMIX and linearly decomposed critics in stochastic TAPE and DOP is also three-layer feed-forward neural network with hidden size 32, where the coefficients for local $Q$ values are always non-negative. Hyperparameter $p$ for ER based agent topology in stochastic TAPE is $0.7$. All algorithms are trained with Adam optimizer \cite{kingma2014adam} and the learning rate is $1\times10^{-3}$. And we use batch size 32 for QMIX. Since we wish to compare differences of the policy gradients, we omit the off-policy target for critic in Eq. \ref{dop_off_target} in stochastic TAPE and DOP for simplicity.

\subsection*{E.2$\quad$Level-Based Foraging}\begin{figure*}[ht]
    \centering
    \includegraphics[width=.9\textwidth]{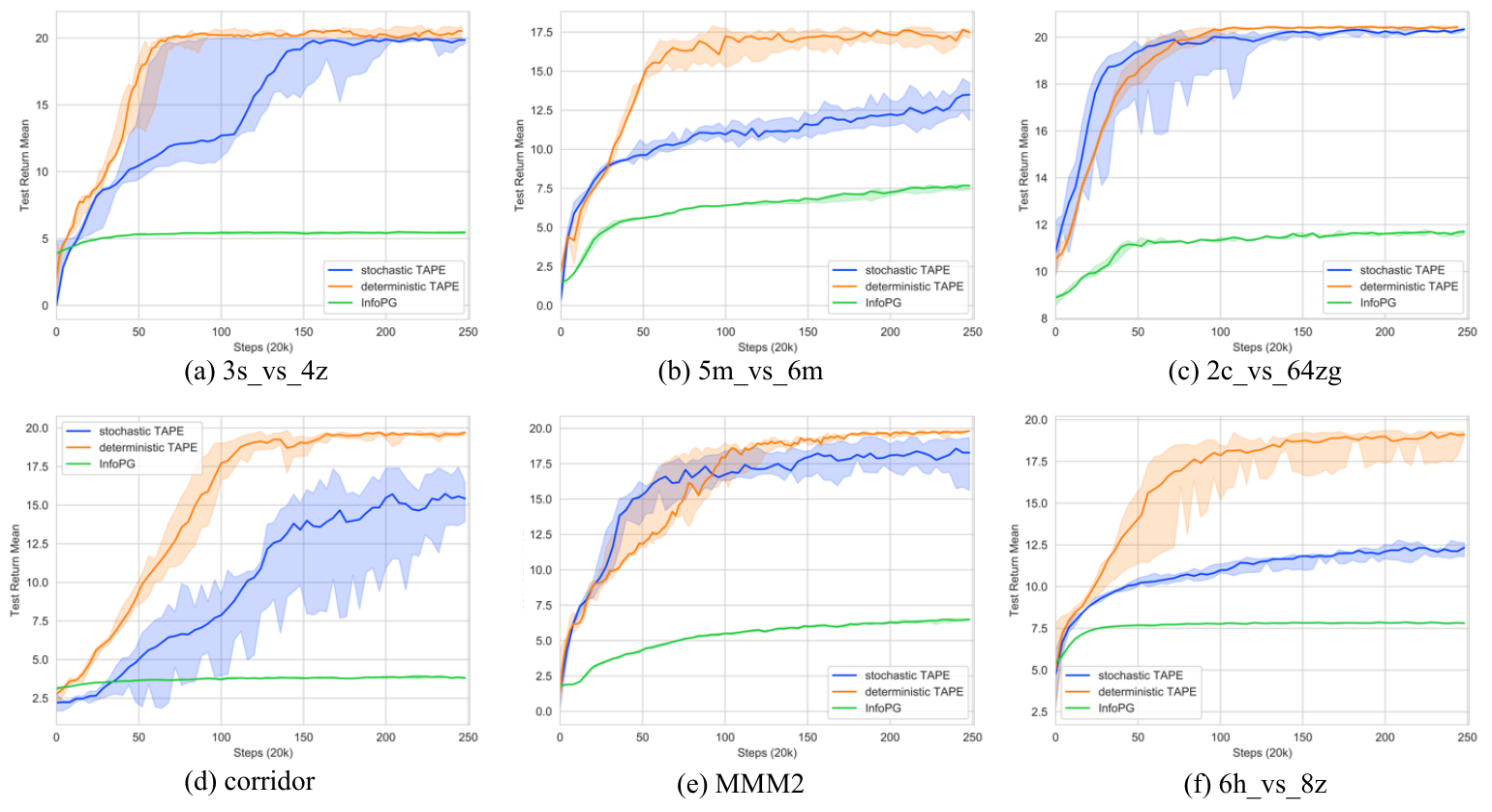}
    \caption{Evaluation results on the six tasks of SMAC. InfoPG is a decentralized graph-based MARL method with networked agents. Our proposed methods adopt agent topology and follow centralized-training-decentralized-execution paradigm, which make them outperform InfoPG by a large margin during decentralized execution.}
\end{figure*}
We use the official implementation of Level-Based Foraging \cite{papoudakis2021benchmarking} (LBF) and remain the default settings of the environment, e.g. reward function and randomly scattered food items and agents. The time limit for 8x8-2p-3f is 25, and 120 for 15x15-4p-5f. For the 8x8-2p-3f scenario, we use the '-coop' option in the environment to force the agents collect all food items together and make the task more difficult. The training goes on for 2 million timesteps in 8x8-2p-3f and 5 million timesteps in 15x15-4p-5f. Each algorithm runs for 100 episodes for test every 50k timesteps. The evaluation metric is the average return of the test episodes.

We use official implementations for all baseline algorithms and implement our methods based on the official implementations without changing the default hyperparameters. For example, there are two hidden layers with hidden size being 64 in the hyper network in QMIX, and target networks are updated every $600$ steps in DOP. For fair comparison, we only change the number of parallel-running environments to 4 for all algorithms and remain the other hyperparameters recommended by the official implementations. Hyperparameter $p$ for both stochastic and deterministic TAPE is $0.3$.

\subsection*{E.3$\quad$Starcraft Multi-Agent Challenge} 
Agents and players do not play the whole StarCraft II game in Starcraft Multi-Agent Challenge (SMAC). In SMAC, a series of decentralized micromanagement tasks in StarCraft II are proposed to test a group of cooperative agents. The agents must cooperate to fight against enemy units controlled by StarCraft II built-in game AI. We keep the default settings of SMAC in our experiments, such as game AI difficulty 7, observation range and unit health point. We run 32 test episodes every 20k timesteps for evaluation and report median test win rate across all individual runs as recommended.

As in LBF, we use the official implementations, recommended hyperparameters and 4 parallel-running environments for all baseline algorithms. The hyperparameter $p$ controlling the probability of edges being present is chosen from $\{0.1,0.3\}$ for stochastic TAPE and $\{0.5,0.7\}$ for deterministic TAPE. $p$ for each map and algorithm is given in Table \ref{hyper_p}.
\begin{table}[h]
    \centering
	\begin{tabular}{c|c|c}
        \toprule
        Map & $p$ (stochastic) & $p$ (deterministic) \\
        \hline

          5m\_vs\_6m & 0.1 & 0.7 \\
          2c\_vs\_64zg & 0.1 & 0.7 \\
        6h\_vs\_8z & 0.1 & 0.5\\
         3s\_vs\_4z & 0.3 & 0.5 \\
          corridor & 0.3 & 0.5 \\
          MMM2 & 0.3 & 0.5 \\

        \bottomrule
	\end{tabular}
     \vspace{0.5em}
    \caption{Hyperparameter $p$ in each map and algorithm.}
    \label{hyper_p}
    
\end{table}
\section*{F$\quad$Additional Experiments}
The concept of topology is also adopted in fully-decentralized MARL methods with networked agents \cite{zhang2018fully,konan2022iterated}. In fully-decentralized methods, the CDM issue does not exist since all agents are trained independently. Agents are networked together according to the topology, so that they can consider each other during decision making to better cooperate and even achieve local consensus. However, although using agent topology to gather and utilize local information of neighboring agents, this decentralized training paradigm cannot coordinate agents' behavior as well as our methods, which is based on centralized-training-decentralized-execution paradigm.

InfoPG \cite{konan2022iterated} is the state-of-the-art fully-decentralized method with networked agents, where agents' policies are conditional on the
policies of its neighboring teammates. We run InfoPG on all six maps of our SMAC experiments and compare the mean test return with our proposed stochastic TAPE and deterministic TAPE.

The results demonstrate that our agent topology is effective in facilitating cooperation and filtering out bad influence from other agents during centralized training, which makes it outperform the fully-decentralized method InfoPG.
\end{document}